# Local Measurement of Current Density by Magneto-Optical Current Reconstruction in Normally and Overpressure Processed Bi-2223 Tapes

S. Patnaik, D.M. Feldmann, A. Polyanskii, Y. Yuan, J. Jiang, X. Y. Cai, E. E. Hellstrom,
D.C. Larbalestier, and Y. Huang

*Abstract*—Magneto-optical current reconstruction has been used for detailed analysis of the local critical current density ($J_c$) variation in monocore Bi-2223 tapes. We find, even in high quality tapes with bulk transport $J_c$ ~40 kA/cm$^2$ (77K, 0T), that there exist local regions which possess current densities of more than 200 kA/cm$^2$. Overpressure processing at 148 bar significantly improved $J_c$ to 48 kA/cm$^2$ by improving the connectivity. For the overpressure-processed sample we find that the current distribution is more uniform and that the maximum local current density at 77 K is increased almost to 300 kA/cm$^2$.

## I. Introduction

AT present, silver sheathed Bi-2223 is the only high $T_c$ superconducting material that can be fabricated in long lengths suitable for large-scale engineering applications[1]. While the best available long length monocore tapes have critical current density $J_c(77K,0T)$ ~ 40 kA/cm$^2$, there is increasing evidence that this is by no means the ultimate limiting value[2,3]. For example, $J_c(77K,0T)$ as high as 70 kA/cm$^2$ has been achieved for short multifilament samples[4]. Poor inter-grain connectivity is one reason advanced for the less than optimum current carrying capacity[5], but prior magneto-optical reconstructions suggest that problems with connectivity have many sources. For example, even very high-$J_c$ multifilament Bi-2223 tapes can contain as much as 20% porosity[6]. Such pores, cracks and other detrimental microstructural features are at least partially caused by the multi-step thermomechanical heat treatment needed to react and to densify the starting Bi-2212 precursor as it reacts towards a largely Bi-2223 phase assemblage. One way to minimize porosity is to heat treat the tape with an overpressure (OP), in which a high pressure, inert gas is used to apply pressure during heat treatment to improve density of Bi-2223[7]. One goal of OP processing is to achieve high $J_c$ tapes without the need for the intermediate rolling step, which is an important cause of residual cracking and poor connectivity, even though the IR step produces significant densification[6].

In this study, we used magneto-optical (MO) imaging to compare distributions of the local critical current density on scales of a few square micrometers in OP and normally processed (NP) samples. Our study indicates that overpressure processing not only increases the magnitude of maximum local $J_c$, but that it also improves overall connectivity,

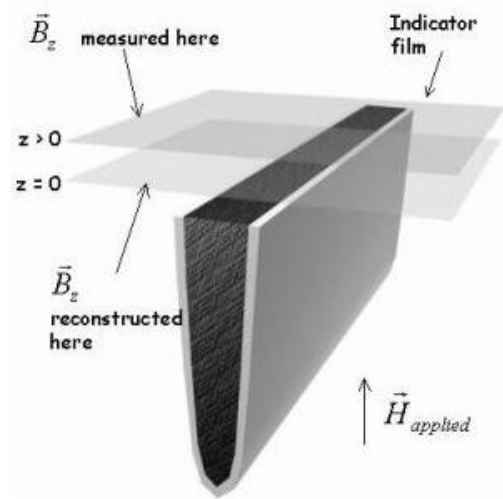

Fig. 1. The slab geometry used for the MO imaging. The specimen is half of a 2 mm wide tape, embedded in a polished plastic puck on which the MO indicator film is laid. The Bi-2223 thickness is 100μm. The external field is applied parallel to the tape plane, that is, approximately parallel to the ab planes.

Manuscript received August 6, 2002. This work was supported by the DOE-EERE, with partial facilities support by the NSF funded MRSEC on Nanostructured Materials and Interfaces.

S. Patnaik, D. M. Feldmann, A. A. Polyanskii, Y. Yuan, J. Jiang, X. Y. Cai, E. E. Hellstrom, and D. C. Larbalestier are with the Applied Superconductivity Center, University of Wisconsin–Madison, WI 53706, USA (phone:608-262-0711;fax: 608-262-1087; e-mail: spatnaik@facstaff.wisc.edu).

Y. Huang is with American Superconductor Copr., Westborough, MA 01581. (phone 508-836-4200, email Yhuang@amsuper.com)

transport critical current density, and the frequency of the high current carrying regions.

## II. EXPERIMENTAL DETAILS

This study is primarily based on two differently processed samples taken from the same monocore silver sheathed Bi-2223 tape. The tape was made by the usual oxide-powder-in-tube (OPIT) process[7,10]. Each sample had the identical first heat treatment at 1 bar pressure (HT1) followed by intermediate rolling. The only difference between the samples occurred in the second heat treatment (HT2). While HT2 for the OP sample was done at 148 bar, that for the NP sample was carried out at 1 bar pressure. The oxygen partial pressure in both cases was 0.075 bar. The details of sample preparation are published elsewhere[7]. The onset $T_c$ of the samples as determined by SQUID measurement was ~ 111 K. The self-field transport $J_c$ evaluated at 1 μV/cm at 77 K was measured to be 48 kA/cm$^2$ for the OP processed and 39 kA/cm$^2$ for the NP processed sample.

Both samples were mounted in a single plastic mount and imaged at the same time. The sample puck was ground to the half width of the tapes. The final polish was with 0.05 μm Al$_2$O$_3$ powder in methanol so as to achieve the best possible flat surface to permit close proximity to the indicator film. The magnetic field maps were taken after cooling the samples in zero field to 77 K and then applying a magnetic field of 46 mT.

The magneto-optic imaging technique is based on the Faraday effect, that is rotation of the plane of polarization of light in the presence of a magnetic field. A Bi-doped iron garnet indicator film which exhibits the Faraday effect is placed on top of the superconducting sample. Polarized light reflected from a mirror coating applied to the underside of the indicator film carries information about the local variation of magnetic field produced by currents flowing in the sample. The angle of rotation of the polarization vector is directly proportional to the magnitude of the applied magnetic field over the range 0-80 mT and thus produces a variation of reflected light intensity that is directly proportional to the perpendicular component of magnetic field in the imaging film.

To permit current reconstructions (MO-CR), a carefully polished, longitudinal cross section at the half width of the tape was used. This is a good approximation of a slab geometry to which the external magnetic field was applied parallel to the tape plane with **H ∥ ab**. This BSCCO tape geometry approximates two important conditions; a) 2D current flow and b) quasi-infinite sample in z direction as is shown in figure 1. If $B_z (x,y)$ in the plane z=0 is known, then applying Maxwell's equation, the current density **J** can be derived as;

$$\vec{J} = -\frac{2}{\mu_0} \vec{\nabla} \times \vec{B}_z$$

$$|\vec{J}| = \frac{2}{\mu_0} \sqrt{\left(\frac{\partial B_z}{\partial x}\right)^2 + \left(\frac{\partial B_z}{\partial y}\right)^2}$$

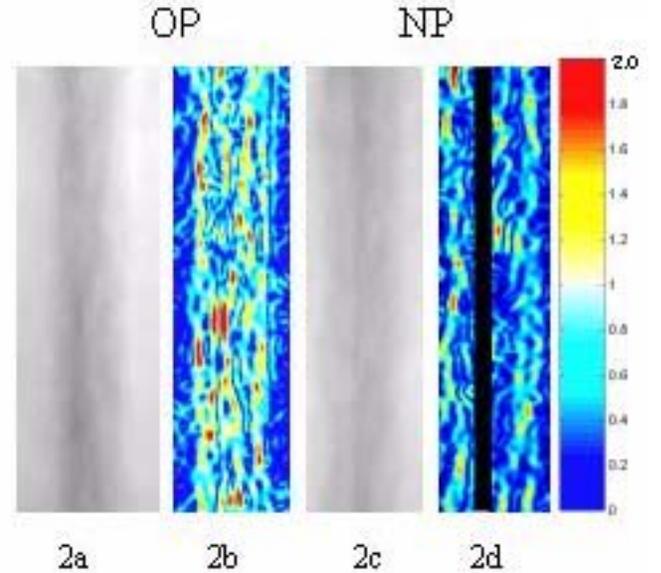

Fig. 2. a) MO image of OP sample, b) MO-CR of OP sample, c) MO image of NP sample, and d) MO-CR of NP sample. The images were taken at T = 77 K, and H = 46 mT. Improvement in local $J_c$ for the OP sample is evident. The color scale of $J_c$ is linear and rises from 0 to 2x10$^5$ A/cm$^2$. The black bar in the NP image blanks out uncertain regions due to erratic domain structure that occurs occasionally in the ferromagnetic film.

Details of the MO-CR algorithm are described by Feldmann[8].

## III. RESULTS AND DISCUSSION

Figure 2 compares the MO image and the MO current reconstruction (MO-CR) for the OP and the NP samples. The sample thickness is ~100 μm and the reconstructions are done over a length of ~500 μm. Both dimensions are much larger than the pixel size (2 x 2 μm) used to represent the data and the effective spatial resolution of the reconstructions which is estimated to be ~5 x 5 μm. Figure 2a and 2b represent the actual MO image and the MO-CR for the OP sample, while Figs. 2c and 2d provide the same information for the NP sample. The reconstructions are color coded to represent the local variability of critical current density. Dark red and dark blue represent the maximum and minimum in local current density for each plot presented.

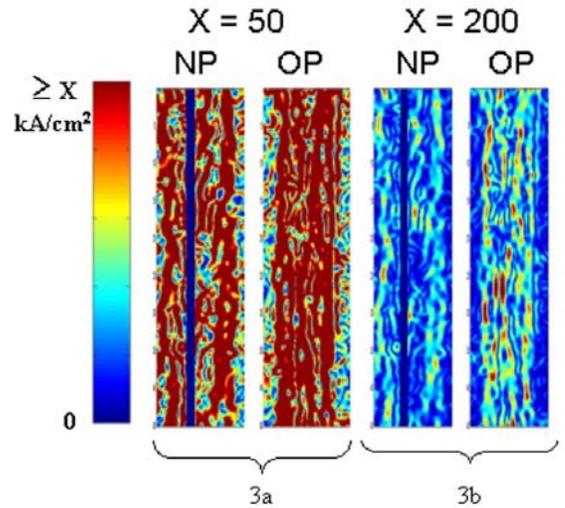

Fig. 3. Direct comparison of connectivity between OP and NP samples; a) for upper limit value (see text) of 50kA/cm$^2$, and b) for upper limit value of 200 kA/cm$^2$. Note that there is a continuous current path for the OP sample (Fig. 3a right) consistent with its transport current density of 48 kA/cm$^2$.

The MO images of figure 2a and 2c represent the non-uniform nature of $B_z$ due to the percolative nature of current flow along the tape's polycrystalline network. This feature is even more evident in the current density maps shown in figure 2b and 2d. It is striking to find that there are local regions in both samples that possess current densities up to 5 times that of the bulk transport current density. The central black line in the NP reconstruction (fig. 2d) blanks out an uncertain region due to erratic domain structure that occurs occasionally in the ferromagnetic indicator film. Such domains were much less pronounced in the OP sample. As is evident from the current reconstructions, we note that the high current carrying regions are much more uniformly distributed in the OP sample. It is also remarkable that high current density regions in the OP sample are much larger in size than in the normally processed sample. This implies improved connectivity in OP sample over multiple grains, each of which is typically ~ 20 μm.

Figure 3a and 3b enable a direct comparison of connectivity between the OP and the NP samples at 77 K using two different visualizations of the MO-CR data. The color scale is always set to go from zero (dark blue) to maximum (dark red), using maxima of X = 50 kA/cm$^2$ and X = 200 kA/cm$^2$ in the two cases presented. Recall that the transport current densities for the OP and NP samples are 48 kA/cm$^2$ and 39 kA/cm$^2$ and that the field suppression of $J_c$ in the condition of H parallel to the tape plane is less than 5%. It is evident that the high current carrying regions in the OP sample vastly outnumber those in the NP sample for both X = 50 kA/cm$^2$ and X = 200 kA/cm$^2$. Further, for X= 50 kA/cm$^2$, we find that almost the whole OP sample is red in color and we note that 50 kA/cm$^2$ is very close to the 48 kA/cm$^2$ self field transport $J_c$(77K).

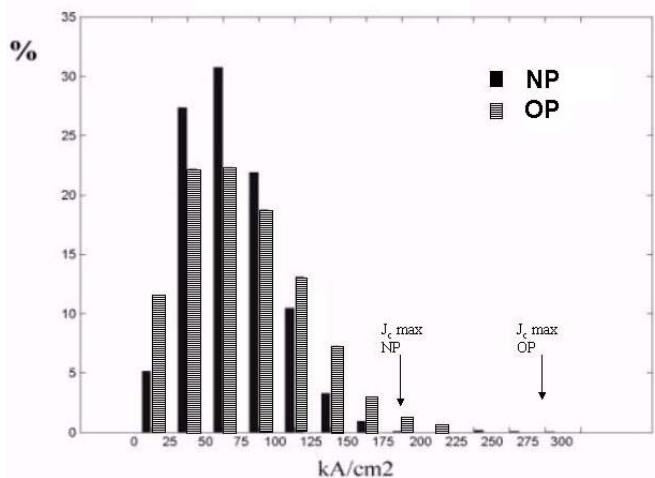

Fig. 4. Histogram of the percentage of local regions that carry a specific critical current density. The average $J_c$ and the local $J_c$ maxima are drastically improved by overpressure processing.

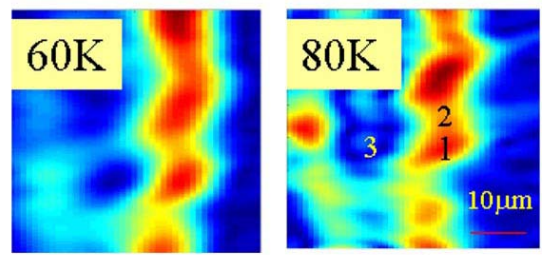
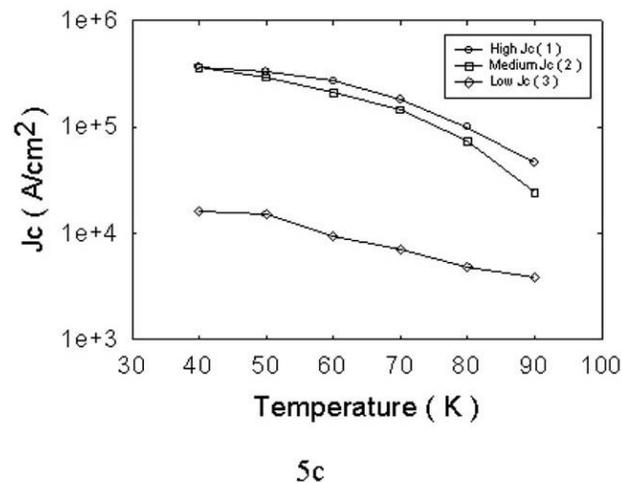

Fig. 5. a) Local $J_c$ at T = 60 K, H = 31 mT for NP sample; b) local $J_c$ at the same region at T = 80K, H = 31mT, spot 1, 2, and 3 correspond to high, medium, and low $J_c$ regions, respectively; c) temperature dependence of local $J_c$ at constant external field. Notice that $J_c$ at spot 2 does not go to 0 at 90 K implying the suppressed $J_c$ is not due to Bi-2212 intergrowths.

To obtain a quantitative estimate of the percentage of high current carrying regions, we show a histogram of the data derived from the reconstructions in figure 4. Each bar corresponds to a range of $J_c$ value between x kA/cm$^2$ and x+25 kA/cm$^2$. From the histogram, it is unambiguously evident that overpressure processing not only increases the density of local regions carrying higher $J_c$ as compared to the NP sample, but also improves the maximum local $J_c$ in the sample. While the maximum observed value for the NP sample is ~ 200 kA/cm$^2$, the OP sample possesses values of nearly 300 kA/cm$^2$ at 77 K. This is at least 30% higher than the previously reported $J_c$ value for the best quality Bi-2223 tapes[9]. We believe that the higher values of the OP sample represent a real improvement associated with closing up of voids, cracks and other obstacles that remain in NP processed tape. It is not yet clear what the ultimate limit to $J_c$ will be for fully optimized OP processed tape[7].

Figure 5 presents another view of how to use MO-CR to investigate the properties of BSCCO tapes. In this case, a study of what obstructs current flow across two neighboring regions that differ in current density by about an order of magnitude is studied. Recent work has shown that 2212 phase is one such obstacle[10, 11]. To test this hypothesis, we have done MO reconstructions as a function of temperature on an NP sample. Figure 5a and 5b represent reconstructions in a small region at 60 K and 80 K respectively. The high $J_c$ region is marked 1, the medium $J_c$ region is marked 2 (the link between high current carrying regions), and the low-$J_c$ region is marked 3. Figure 5c shows the temperature dependence of local $J_c$ in these 3 regions. Even at 90K region 2 of medium $J_c$ magnitude does not drop to zero, therefore we conclude that Bi-2212 intergrowths are not solely responsible for poor connectivity at these regions in this Bi-2223 tape.

## IV. CONCLUSIONS

We have done extensive magneto-optical current reconstructions on Bi-2223 tapes, which were processed under 1 bar and under 148 bar pressure. We have established that overpressure processing significantly improves the local critical current density, leading to better connected, higher $J_c$ regions and higher maximum $J_c$ values. We have observed values of $J_c$ of almost 300kA/cm$^2$ in the overpressure processed tape. Our study shows that connectivity is a multi parameter function and Bi-2212 regions or intergrowths are not the sole interruption to current flow.

## ACKNOWLEDGMENT

We wish to thank colleagues in the Wire Development Group for many useful discussions of these results.